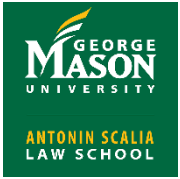
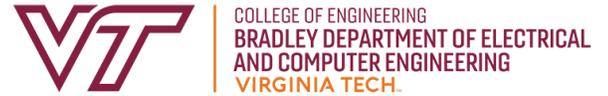

# Cyber Threats Jurisdiction and Authority – A Legal Discussion

Feras A. Batarseh

Antonin Scalia Law School, George Mason University

Bradley Department of Electrical and Computer Engineering, Virginia Tech

batarseh@vt.edu

**Abstract**: Cybersecurity threats affect all aspects of society; critical infrastructures (such as networks, corporate systems, water supply systems, and intelligent transportation systems) are especially prone to attacks and can have tangible negative consequences on society. However, these critical cyber systems are generally governed by multiple jurisdictions, for instance the Metro in the Washington, D.C. area is managed by the states of Virginia and Maryland, as well as the District of Columbia (DC) through Washington Metropolitan Area Transit Authority (WMATA). Additionally, the water treatment infrastructure managed by DC Water consists of waste water input from Fairfax and Arlington counties, and the district (i.e. DC). Additionally, cyber attacks usually launch from unknown sources, through unknown switches and servers, and end up at the destination without much knowledge on their source or path. Certain infrastructures are shared amongst multiple countries, another idiosyncrasy that exacerbates the issue of governance. This law paper however, is not concerned with the general governance of these infrastructures, rather with the ambiguity in the relevant laws or doctrines about which authority would prevail in the context of a cyber threat or a cyber-attack, with a focus on federal vs. state issues, international law involvement, federal preemption, technical aspects that could affect lawmaking, and conflicting responsibilities in cases of cyber crime. A legal analysis of previous cases is presented, as well as an extended discussion addressing different sides of the argument.

## 1. Introduction

One of the most important constitutional doctrines is the Supremacy Clause, which is Article VI; Paragraph 2 of the U.S. Constitution[1]. It establishes that the U.S. Constitution and federal law take precedence over state law. However, in many cases that span state or even national borders (such as issues of international law), the Supremacy Clause can be a target of multiple interpretations and different opinions, which lead to different court cases' outcomes.

Cybersecurity is a field of law that requires considering the Supremacy Clause however, legal scholars should reflect on how it is to be applied and in what form given the field's technical, virtual, and non-geographic nature. This report aims to present different factors and aspects that influence that discussion.

The focus of the report is on U.S. law; however an example from Australia (that referenced U.S. statutes in their court proceedings) illustrates the need for such a discussion across multiple jurisdictions. In the year 2000, the Maroochy Water Services was compromised via a cyber attack that paralyzed the water plant, dumped one million liters of sewage into the river, and -more dangerously- polluted 500 meters of open drain in a residential area. In an opinion published by the Supreme Court of Queensland[2], the court concluded that if the plant had applied the U.S. National Institute for Standards and Technology (NIST)'s SP 800-53 standard[3] (for Security and Privacy Controls for Information Systems and Organizations), the threat would have been mitigated or its effects hugely minimized. In the U.S., similar threats have occurred (and continue to), such as a recent attack on Florida's water system in 2021, and other wide-scale attacks across the nation on critical infrastructure such as the infamous multi-state Colonial gas pipeline attack. In some cyber attack cases, the jurisdiction is clear because all the components of an infrastructure (such as a water plant) are under one district. However, in most other U.S. cases presented in this report (that span cross states or international borders), the cyber attack effects extend to multiple areas, which creates the need for further legal analysis – this law paper aims to discuss that.

The remaining of this report is presented as follows: the next section (2) introduces the legal questions addressed. Section 3 presents four arguments related to the questions under discussion: (1) federal law argument, (2) state law argument, (3) the technical argument, and (4) the international law argument. Lastly, section 4 presents final remarks and conclusions.

## 2. The Legal Question(s)

The question that this report addresses is three-fold: (1) in cases of cyber attacks on shared infrastructures or networks, which jurisdiction should prevail? (2) What precedential and constitutional considerations are to be debated to deliver legal answers related to cyber crimes? (3) How does federal vs. state vs. international doctrines apply in cases of Cybersecurity law?

### 3. Evaluation (The *Four* Main Arguments)

This section provides a discussion on both sides of the argument, i.e. what points support the need for the federal government's oversight versus the need for state's involvement and in some cases, international law considerations. Section 3.1 makes the case for federal leadership in this arena, which 3.2 looks at state-driven use cases of Cybersecurity, section 3.3 reviews the technical aspects related to the law and lawmaking, and lastly, section 3.4 addresses the international law angle and its relevance to the jurisdiction discussion.

*3.1 The Federal Argument: Authority Granted through FISA and the Executive Branch*

In many cyber-related instances, the federal government's intervention is agreed on and needed, besides protecting consumers on a national scale; but also when protecting American networks and infrastructure against external attacks by state and non-state players. In many cases, cyber attacks require a response on a national scale, such as in the cases of Solar Winds and other similar adversaries. Article I, Section 8, Clause 11 of the U.S. Constitution grants Congress the power to declare war (including a cyber war). The President however, being the Commander in Chief, derives the power to order the attack after a congressional declaration. These provisions dictate coordination between the president (Executive Branch) and Congress (Legislative Branch) on all cyber war affairs.

Additionally, the government has been getting involved in Cybersecurity entanglements internally for a while now. For instance, in a recent case, the Federal Trade Commission (FTC) "alleged that LabMD's data-security program was inadequate and thus constituted an unfair act or practice" under Section 5 (a) of the FTC act, 15 USC § 45 (a) (as in FTC v. LabMD). The FTC act prohibits "unfair or deceptive acts or practices in or affecting commerce". In 2005, the FTC started with actions under the mentioned provision against companies with potentially deficient security of their networks or ones that failed to protect consumer data against hackers. The vast majority of these cases have ended in some form of a settlement – which constitutes an example of a successful federal intervention in cyber matters.

Moreover, in 2008 and 2009, hackers (unknown) successfully accessed Wyndham Worldwide Corporation's computer systems. They stole personal information "for thousands of consumers leading to over $10.6 million dollars in fraudulent charges, the FTC filed suit in federal District Court, alleging that Wyndham's conduct was an unfair practice and that its privacy policy was deceptive" (refer to FTC vs. Wyndham Worldwide).

Similar domestic cyber threat cases seem to need and require the intervention of the federal government to apply generic rules to govern the process, especially that they are considered clear and responsibility could be deemed fairly manageable, nonetheless, in cases where an attack is from an external player (international) via software and data tools, it is difficult to understand responsibility and even qualify the action as an act of war (such as a possible routine surveillance operation for instance). Justice Scalia argued (in one of the cases related to Guantanamo and acts of war) that the judiciary branch has no business (whatsoever) in war making, whether it is in cases of cyber or not.

Nonetheless, in cases related to Foreign Intelligence Surveillance Act (FISA), cyber defense and offense, and other means of technological acts of war (such as drones), it seems that the statute is under scrutiny due to its lack of "catching up" with technology advancements, which is an area that needs progress and further collaboration with other countries.

Similar to Environmental law (where smoke might cross borders and water bodies are shared), cyber attacks cross borders, and enacting centralized/federal laws in isolation seems disconnected, accordingly, a combination between international collaboration while prioritizing U.S. national security is needed in enacting or updating laws such as FISA and other Department of Defense regulations regarding recommended actions in cyber and other advanced domains of war, in addition to the Computer Fraud and Abuse Act (CFAA) which is the one of the main methods to prosecuting "federal level" cyber crime.

It is important to note that the President is allowed (through emergency powers), in cases of rebellion, such as in the famous Prizes cases 67 U.S. 635 to act swiftly (a case argued before the Supreme Court of the United States in 1862 during the American Civil War. The Supreme Court's decision declared the blockade of the Southern ports ordered by President Abraham Lincoln constitutional). Similarly, in cases of a cyber attack on critical infrastructures such as on power grids, gas pipelines, or internet networks, the president can act swiftly and launch a cyber war as a measure to avoid extensive harm to the country.

As Koh[4] discussed in the Emerging Law of 21$^{st}$ Century War, a *temporary dictatorship* (quick, major, and centralized Presidential decisions) could be warranted due to cyber cases and their complexity in emergency situations.

Additionally, in the foundational case, Youngstown v. Sawyer (1952), it provided a solid precedent and path forward for many similar cases where the line might blur between congressional distinct right to declare cyber attacks as acts of war and a president's executive order that enables such acts either based on proactive (speculative) or reactive measures (in the Youngstown case, an action as seizing a steal manufacturing plant due to the nation's involvement in war was invalid), and only vested (by the constitution) to lawmakers.

As granted by the U.S. Constitution, only congress can authorize war, however, the President, being the Commander-in-Chief can order the military to take action in some emergency scenarios without Congress's approval. In Youngstown v. Sawyer, the court took note that the President has notified congress of the actions, and didn't act in secrecy – albeit not a notion that would apply to Cybersecurity law directly. In states of cyber emergency, the Executive branch can act without Congress (such as in the Prize cases, 67 U.S. 635). It is important to note that executive orders by the US president, such as orders 13636 (Improving Critical Infrastructure Cybersecurity in 2013) and 14028 (Improving the Nations Cybersecurity in 2021) have major implications on domestic reaction to cyber attacks. The orders mentioned aspects that won't be possible if the federal government is not involved, for instance: (1) "removing Barriers to Sharing Threat Information" between states, as well as business and companies (such as in the case of FireEye after the Solar winds attack), (2) enforcing the role of The Cybersecurity and Infrastructure Security Agency (CISA) as a governing agency, (3) protecting naturally cross state systems such as software supply chains and other critical infrastructure, and (4) other "policy coordination" amongst states, the government and the private sector, and possibly other nations. But how does Stare Decisis present the international dimension?

In famous cases such as U.S. citizens in Nicaragua vs. Reagan, it was held that statutes supersede customary international law (discussed in section 3.4) and are not subject to challenge on the basis of a violation of international law, which leads to having exits when it comes to acting solely in case of a cyber attack. However, can the US really act alone in cases of attacks? What about international collaboration and cyber threats that cross borders?

An action that the president could take in cases of a cyber attack that spans multiple countries is notifying the UN's Security Council and getting approval to have the response qualified as self-defense. Conversely, if for instance, state A is a victim of a cyber attack through an entry point in state B, which jurisdiction prevails, A, B, or both? The mentioned executive orders and federal laws continue to play a bipartisan major role in driving the issue at the federal level, in hopes that it would trickle down to the state level, which is discussed in the next section. The next section presents discussions related to that and the state's law argument for governing cyber threats.

*3.2 The State's Argument: Authority Granted through the Constitution*

As it is established prior in this report, cyber issues are not governed by geographical state borders or other definitions such as by districts and district courts. Nonetheless, there have been cases where state or district courts dealt with a cyber attack albeit from a consumer protection angle.

For instance, in 2013, in a non-federal case, hackers attacked Neiman Marcus (the department store), and stole the credit card numbers of its customers. In December of the same year, the store was informed by some of its customers about fraudulent charges on their credit cards. In January, it announced to the public that the cyber attack had occurred and that 350k cards had been exposed to the hackers' malware. In the wake of those disclosures, several customers brought this action under the Class Action Fairness Act[5], 28 U.S.C. § 1332, seeking various forms of making up for data breaches and losses. The district court stopped the suit in its tracks; however, the court ruled that both the individual plaintiffs and the consumer group lacked standing under Article 3 of the US Constitution. This resulted in a dismissal of the complaint without prejudice (see Hilary Remijas vs. Neiman Marcus).

Conversely, in another non-federal case, Plaintiffs Luciano Pisciotta and Daniel Mills brought an action on behalf of an alleged group of consumers and potential customers of the Old National Bancorp (ONB) – one of the largest financial services holding company headquartered in the state of Indiana. The case was brought to court after ONB's website solicited personal information for banking services and failed to secure it properly. Accordingly, a third party hacker was able to collect confidential information of multiple thousands of ONB customers. The plaintiffs sought "damages relief".

These two example cases (ONB and Neiman Marcus) show how states and district courts deal with cases related to Cybersecurity. However, the main issue with state laws relevant to cyber-issues is that they have different "technical" assumptions that lead to different outcomes/decisions while crafting the laws. States have different laws for Cybersecurity, and where one lives certainly effects what laws are in place especially in cases when an attack occurs. For instance, the National Conference of State Legislatures (NCSL) compiled/announced the different laws governing states across the country, Figure 1 illustrates how protected a citizen is by state laws "strictness" as analyzed by the NCSL[6] (states such as TX and AL have stricter Cybersecurity laws while others such as Mississippi and Kentucky have less strict laws).

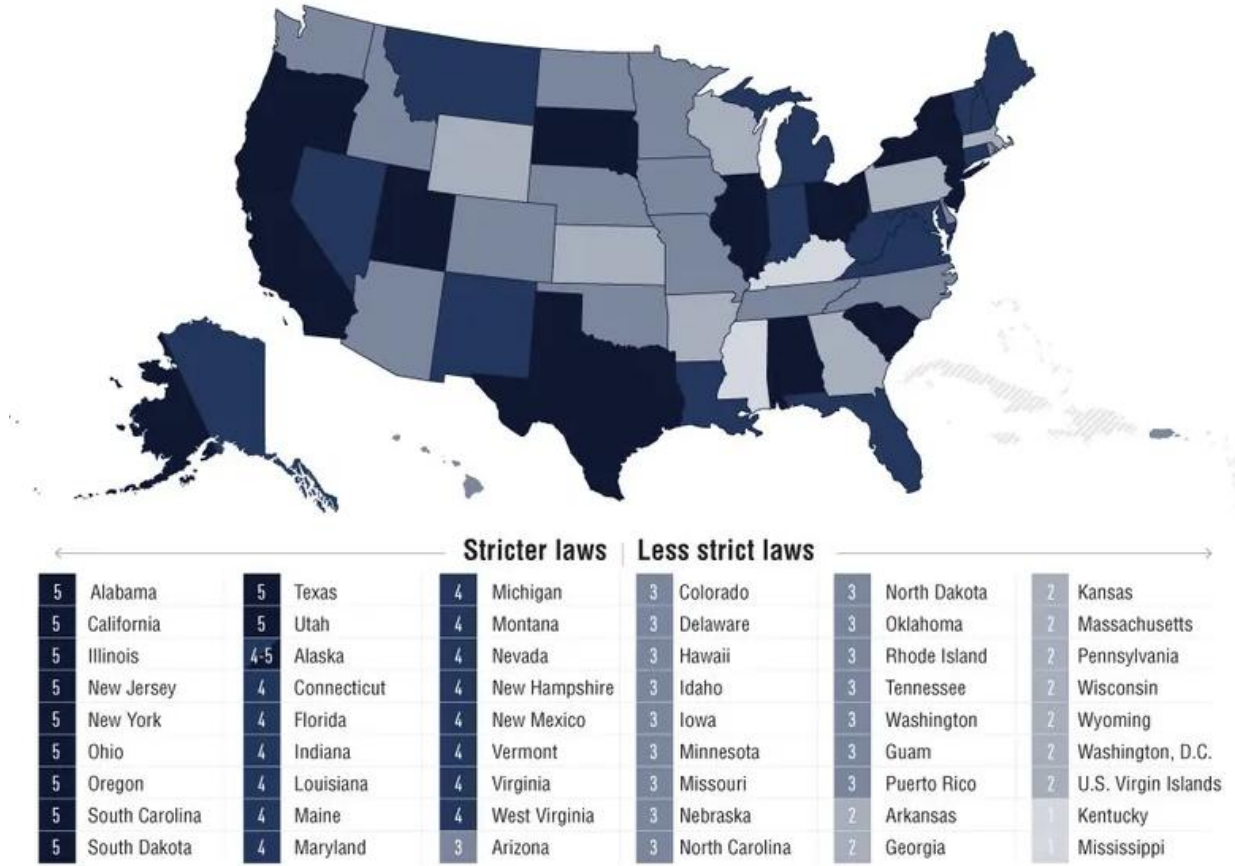

Figure 1: Cybersecurity State Law Strictness (Small Biz Trends[6])

Besides the physical/geographical challenge, there is a technical challenge. In many court cases, responsibility has been proven or disproven based on technical expert opinions, or existing patents related to Cybersecurity, however, even those two foundational sources have been in debate, for instance, when it comes to Cybersecurity inventions or patents, a bedrock principle of patent law is that the claims of a patent define the invention to which the patentee is entitled the right to exclude; Phillips, 415 F.3d at 1312 (quoting Innova/Pure Water, Inc. v. Safari Water Filtration Sys., Inc., 381 F.3d 1111, 1115, Fed. Cir, year 2004). For such detailed technological specifics, courts first "look to the words of the claims themselves to define the scope of the patented invention"[7] (Vitronics Corp. v. Conceptronic, Inc., 90 F.3d 1576, 1582). The claim terms are "generally given their ordinary and customary meaning" but "a patentee may choose to be his own lexicographer and use terms in a manner other than their ordinary meaning". To that point, court cases related to cyber have been affected by the definition of the technology, or type of attack (such as in: Cupp Cybersecurity vs. Trend Micro Inc., United States District Court, N.D. Texas, Dallas Division, 2021). Accordingly, the technical angle and argument is certainly of relevance and is presented in the next section.

*3.3 The Technical Argument (Why Lawmaking Should Catch-up with Technology)*

This section presents an important angle to the discussion; because not all cyber attacks are created equal, and while some attacks are easily and obviously classified as intentional "attacks", others could be less of an attack and more of a data privacy issue, which is not adversarial per se, albeit still unconstitutional. For instance, in the last ten years, the Supreme Court decided several cases on cyber and technological "search and seizures". In USA v. Jones for example, the court decided (year 2012) that GPS trackers usage to monitor the location of a car is a search that is under the umbrella of the 4th amendment of the constitution. In this case, the police installed GPS monitoring tool(s) in a car to monitor the driver for a month. All nine justices of the Supreme Court agreed that these actions constituted a breach to the constitutional rights of the driver.

Debating jurisdiction without the technical components is certainly missing a major component that would influence the effectiveness of crafted Cybersecurity laws[8]. As Figure 2 illustrates, not all cyber threats are created equal, which poses the question: does that have a connection to which jurisdiction prevails in cyber scenarios? For instance, if an attack of some kind is generated through proxy servers in another three or four state, which jurisdiction would prevail? Also, if the attacker was identified, and it's a competitor that is based in another state, which state's law prevails? In common law, the plaintiff "generally" has a choice to "where" to sue, however, in cases where the location is not known such as in "black-box attacks"[9], how should jurisdiction be decided?

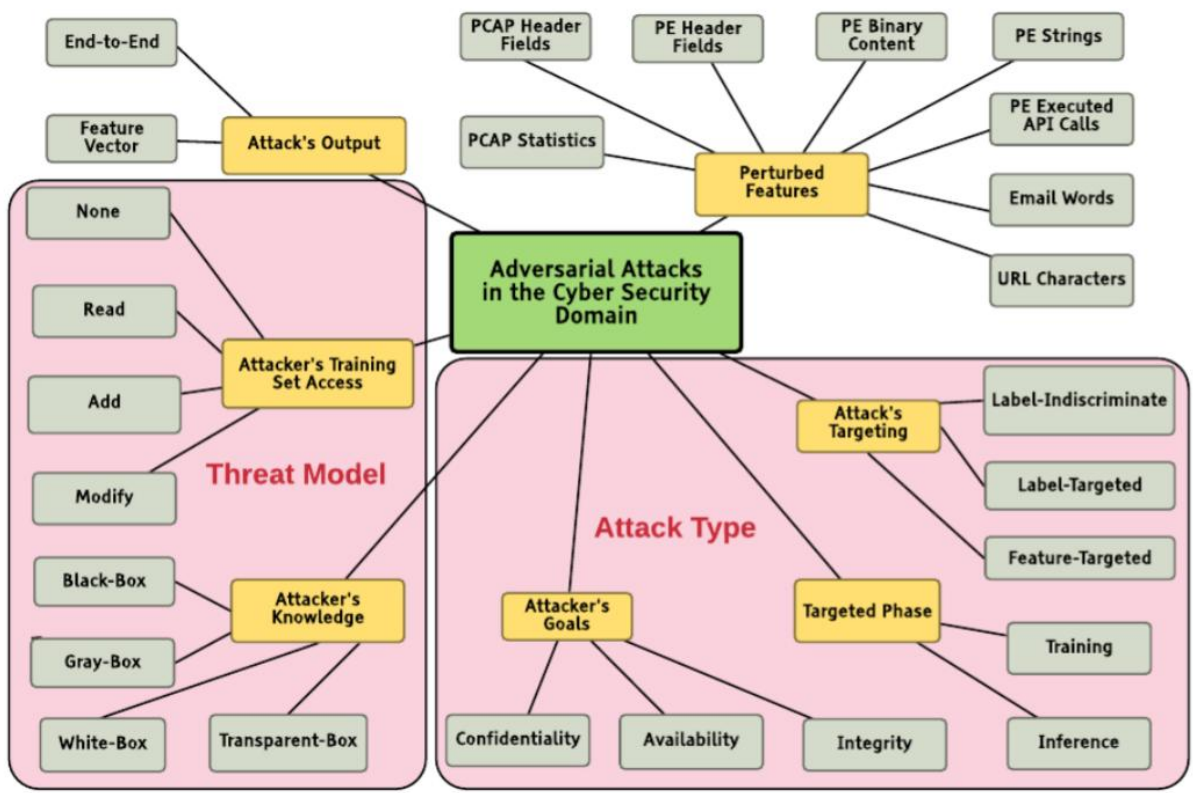

Figure 2: Not All Cyber Threats are Created Equal[9] (Rosenberg, 2020)

More dangerously, when cyber attacks are performed across international borders, such as cases by waged attacks by China, Russia, and Iran on the US (Alexander and Jaffer, 2020)[10], usually they occur through a hidden pattern to hide the country of origin, which leads to cases where the country of origin is either known or unknown. As Alexander and Jaffer (2020)[10] declare: "we predicted Iran would continue to press its advantage and test American resolve. It has done just that . . . over the last decade, Iran has repeatedly hit the U.S. and our allies with both disruptive and destructive cyber attacks"[10]; an ongoing issue that needs clear legal grounds; such as it was challenged by the Huawei vs. United States case, where Huawei challenged the constitutionality of the ban by the US on Huawei to trade or deal with American companies.

The Huawei case was driven by technical observations, and detecting their intervention was only possible due to the technical ability of American investigators and Cybersecurity engineers. From a Constitutional side though, Article I, Section 9, Clause 3 of the Constitution (the Bill of Attainder Clause) clearly states: "No Bill of Attainder or ex post facto Law shall be passed". According to the Supreme Court, that is "a legislative act which inflicts punishment without a judicial trial".

More importantly, the *Bill of Attainder Clause* doesn't mean that the congress is not able to pass a ban on specific firms (such as Huawei) in the event that it their practices are deemed a threat to national security (as discussed in the federal argument section). In such cases (as well as cases similar to the *Solar winds* and *Holiday bear* attacks), the US has to react in ways that doesn't break domestic or international law but protects our homeland, the international dilemma (of Cybersecurity responsibility and governance) is discussed in the next section.

### 3.4 The International Law Point of View (and Argument)

Besides works by the United Nations Commission on Science and Technology for Development (CSTD), one of the most prominent documents calling for international cooperation on Cybersecurity is the Tallinn manual[11]. Many instances of the need for cyber coalitions are reported[12, 13, 14, 15], and there have been many attempts at regulating an international framework for cyber through the management of data and AI goals (some of which are Listed in Table 1).

Table 1: Federal and International Regulation Attempts for Governing Data, Cyber, and AI

| Title and agency | Publication description | Details | Publishing date |
|---|---|---|---|
| (DoD/DIU) Responsible AI Guidelines in Practice | Data protection | The Defense Innovation Unit (DIU) launched an important initiative in 3/2020 to implement the Department of Defense's (DoD) main principles for Cyber and AI into commercial prototyping and acquisition programs. The result is a set of guidelines that could be consulted when creating wide-scale government best practices and recommendations related to the security of data systems and AI algorithms across the government, private sector and academia. | 11/15/2021 |

| (HHS/FDA) Good Machine Learning Practice for Medical Device Development: Guiding Principles | Medical data access and cyber threats | The U.S. Food and Drug Administration, the United Kingdom's Medicines and Healthcare Products Regulatory Agency, and Health Canada have jointly identified ten main guiding principles that can inform the development of Good Machine Learning Practice (GMLP). The principles in this document aims to create a comprehensive legal proposal that can help in protecting medical devices, bio-systems, healthcare data, and hospital operations related to Cybersecurity. | 10/27/2021 |
|---|---|---|---|
| (VA) Department of Veterans Affairs Artificial Intelligence (AI) Strategy | Strategy documents | In 2021, the VA implemented a set of internal regulations for governing data and AI deployments with the goal of increased security. | 10/14/2021 |
| (White House) Executive Order Promoting the Use of Trustworthy AI in | High level guidelines | This Executive Order aims to accelerate Federal adoption of secure AI, modernize government, and cultivate public trust in data-driven algorithms and by defining principles for the use of AI in government in a | 12/08/2020 |
| the Federal Government | | secure manner. It also aims at establishing a generic and common set of policies for implementing and enforcing the principles, which includes developing catalogues for Cybersecurity use cases that would lead to trustworthy AI and other decision support systems. Other orders[16, 17] as presented prior, numbers 26645 and 13636. | |
| (FOC) FOC Joint Statement on Artificial Intelligence and Human Rights | Cyber and AI principles; International declaration | The US is a founding member of the 32 country-member Freedom Online Coalition (FOC). The US work with multiple stakeholders from around the world to develop this statement on human rights, with the goal of refraining from the use of digital systems for repressive and authoritarian purposes around the globe. Additionally, the goal was to ensure the design, development, and use of systems at governments is deployed in accordance with international human rights laws and obligations. | 11/05/2020 |
| (ODNI) Artificial Intelligence | AI/data science in the | This AI and security ethics framework for the Intelligence Community (IC) provides a guide | 07/01/2020 |

| Ethics Framework for the Intelligence Community | intelligence community | for IC personnel on how to design, build, use, protect, consume, and manage secure and safe AI and related data systems. This framework is aimed as serving as a living document to provide stakeholders around the country with a law-backed approach to judgment and to assist with the documentation of issues and cases associated with Cybersecurity. | |
|---|---|---|---|
| (DoD) Ethical Principles for Artificial Intelligence | Data/cyber ethical principles | These AI ethical principles were adopted to enhance the department's commitment to upholding the highest ethical standards during the accelerating adoption of AI and data systems. The principles are based on Military law (US Army guidelines) and the Constitution, Title 10 of the U.S. Code, Law of War, as well as current international laws and treaties, and other commonplace technical concepts. | 02/24/2020 |
| (OECD) Recommendation on AI | International document by the OECD | The Organization for Economic Co-operation and Development (OECD) Recommendations on AI were adopted in mid-2019 by member | 05/01/2019 |

| | | countries. The document promotes AI and related security practices that are inventive and trustworthy while respecting human rights and international law. The recommendation defines principles for the responsible stewardship of secure and safe AI, along with adopting national policies and international cooperation aspects for implementing these principles at the intersection of AI and Cybersecurity. | |
|---|---|---|---|

Accordingly, as Table 1 presents, different domestic and international regulations are under construction and consideration – the table is not conclusive nonetheless, but the question remains on which treaties would be ratified in the US and could eventually govern the Cybersecurity space nationally. Lastly, it is important to note that if the argument #3 (technical issues and crossing geographical boundaries) and argument #4 (international considerations) are unified; it becomes evident that an international take on the law driven by multi-nation collaborations is needed to better mitigate adversarial issues, stop unfriendly states from launching or turning a blind eye to attacks from their lands, and minimize the overall effects of cyber threats around the world.

## 4. Final Remarks and Conclusions

As this report illustrated, in the case of cyber threats, although jurisdiction is debatable, coordination (amongst states and countries) seems to be a must. Not all attacks are created equal (and not all U.S. states have comparable laws), and so the legal treatment should be applied and analyzed differently. As it is stated through FISA principles, *jus bello* ought to be considered as a doctrine in cyber situations just like they are in conventional war. *Jus ad bellum,* as it refers to the conditions under which countries such as the U.S. may utilize to launch an attack or counter attack i.e. has to be consistent with international law, United Nations Charter of 1945, and other treaties such as through NATO. While *jus in bello* regulates the conduct of parties engaged in conflicts, *jus ad bellum* manages the conditions and whether involvement is permissible to start with. The issue with these two concepts is that in cyber situations, quick action might be required (as illustrated the Presidential intervention discussion in section 3.1). For instance, technologies such as automated data-driven cyber attacks (often referred to as Counter Artificial Intelligence; C-AI) could launch attacks reactively without consulting an operator, or a human. Therefore, an ongoing unresolved legal dilemma still persists, such as: who is responsible for loss and damages? And which jurisdiction (again) shall prevail.

Moreover, the issue exacerbates as a Cyber Supply Chain Risk Management Processes (CSCRM) are not in place, as NIST defines it: "the process of identifying, assessing, and mitigating the risks associated with the distributed and interconnected nature of IT product and service supply chains", another notion that needs technical answers to drive the crafting of effective Cybersecurity laws and understanding jurisdiction assignment correctly.

As the number of attacks is on rise (nationally), the rising trend is predicted to sustain[18] (as illustrated in Figure 3). Therefore, this question (jurisdiction assignment) which is at the intersection of technology and law is certainly worthy of further legal analysis and study.

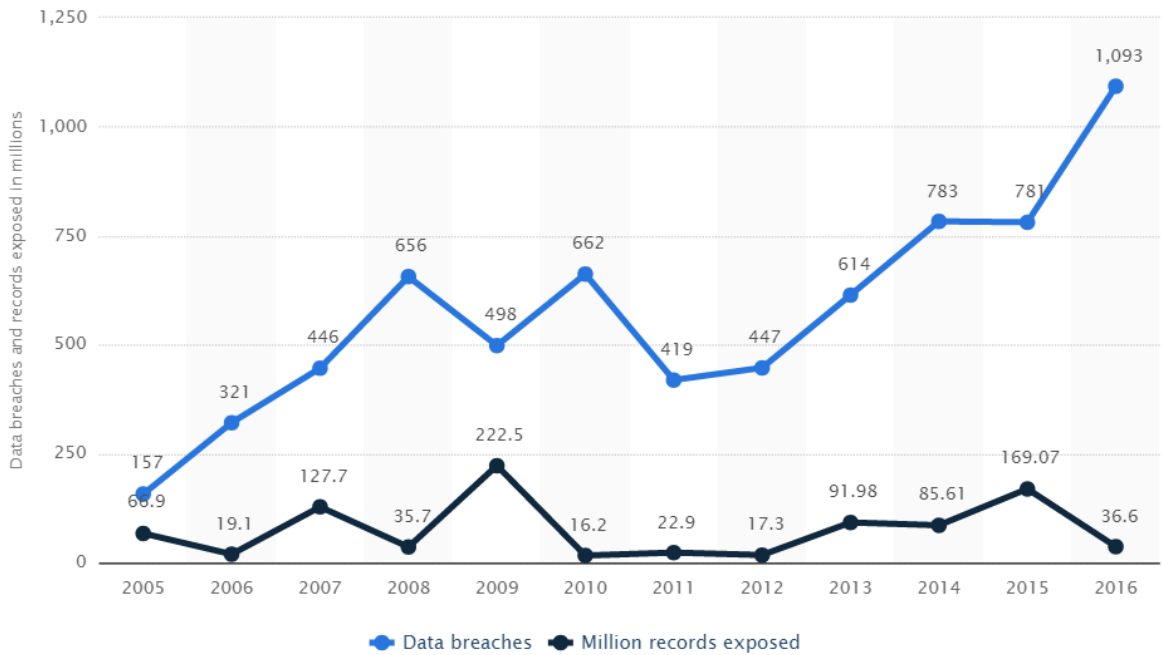

Figure 3: Cybersecurity attacks-induced loss of American records [18]

This law report reviewed presented some pointers to addressing the jurisdiction question from multiple perspectives; however, although more cases are becoming a part of the legal body of knowledge, it is clear that Cybersecurity law is still a fairly growing area that shall be inspected in light of future shifts in technology and international collaborations while preserving US law and Constitution.